# Polarization control in Nitride Quantum Well Light Emitters Enabled by Bottom Tunnel-junctions

Henryk Turski[1,2], Shyam Bharadwaj[1], Huili (Grace) Xing[1,3,4], Debdeep Jena[1,3,4]

[1]*Department of Electrical and Computer Engineering, Cornell University, Ithaca, NY 14853 USA,*

[2]*Institute of High Pressure Physics, Polish Academy of Sciences, Sokołowska 29/37, PL-01-142 Warsaw, Poland*

[3]*Department of Material Science and Engineering, Cornell University, Ithaca, NY 14853 USA,*

[4]*Kavli Institute for Nanoscale Science, Cornell University, Ithaca, NY 14853 USA*

**Abstract.**

The frozen internal polarization-induced electric fields due to broken inversion symmetry in all conventional blue and green nitride semiconductor light emitting semiconductor quantum well heterostructures point in a direction opposite to what is desired for efficient flow of electrons and holes. This state of affairs has persisted because of the desire to have p-type hole injectors on top of the quantum well active region. Because of the internal polarization fields in nitride heterostructures, there exist four permutations of doping and polarization for the realization of such light emitters. Which permutation is the most desirable for efficient light emission? In this work, we answer this question by demonstrating a fundamentally new approach towards efficient light emission with *bottom-tunnel junctions*. The bottom-tunnel junction design aligns the polarization fields in a desired direction in the quantum well, while simultaneously eliminating the need for p-type contacts, and allowing efficient current spreading. By preventing electron overshoot past quantum wells, it disables carrier recombination in undesired regions of the quantized heterostructures, and opens up the possibility for new geometries of integrating and stacking multiple light emitters. Due to the inherent advantages, the bottom-tunnel junction light emitting diode enables a 200-300% increase in the light emission efficiency over alternate heterostructure designs.

**Introduction**

Quantum well semiconductor heterostructures based on the Gallium Nitride (GaN) semiconductor system have made it possible to realize revolutionary photon emitters in the visible and ultraviolet wavelengths [1]. The development of GaN light emitting diodes (LEDs) has enabled semiconductor-based lighting for modern residential and industrial buildings, automobiles, and mobile electronic gadgets, and has also brought lighting to regions of the world that have no direct access to electricity. The high resistivity of p-type layers arising from deep acceptor ionization energies, and undesirable electric field orientations due to the prevailing quantum heterostructure design of these semiconductor materials forces the LEDs to be operated under low current injection conditions to avoid a) losses arising from high-resistance p-type contacts and b) overshoot of hot electrons over the quantum wells causing undesired recombination. These effects become more severe at high current injection and determine the lower limit for threshold of laser diodes (LDs). Since laser-lighting is being investigated as an energy-efficient replacement of LEDs, fresh ideas rooted in the physics of the semiconductor heterostructures are needed to achieve high efficiency photon sources for the future.

One of the most important distinguishing features of III-nitride semiconductor heterostructures is the presence of built-in electronic polarization due to the lack of inversion symmetry in the wurtzite crystal structure [2]. The difference in polarization manifests as the presence of fixed sheet charges of sheet density $\sigma_\pi$ at heterojunctions, which by Gauss' law produce internal electric fields $F_\pi = q\sigma_\pi/\varepsilon_s$, where $q$ is the electron charge and $\varepsilon_s$ is the semiconductor dielectric constant. Typical polarization-induced fields are in the range of $F_\pi$~1MV/cm. Because the electrostatics of p-n junction diodes and the internal quantum efficiency (IQE) of the quantum well regions (due to the quantum-confined Stark effect) are highly sensitive to these electric fields, it is important to explore the various possible orientations and combinations of polarization-induced fields and junction fields that develop due to depletion of carriers.

For vertical LEDs, spontaneous and piezoelectric fields can either enhance or weaken the emission performance of the final device [3]. The direction of the built-in field with respect to the direction of forward current flow in the diode is a critical parameter. Due to the large asymmetry between conductivity of n-type and p-type GaN, almost all vertical p-n diodes are produced using n-type substrates. Lack of highly conductive p-type substrates requires n-type electron injection from the bottom n-type substrate into the quantum well active regions. Being restricted to an n-type substrate, the choice of crystal growth direction – Ga vs N polar – determines whether the polarization field in the quantum wells points in the same or opposite direction to the p-n junction field, and to the direction of current flow.

It has been long believed that higher energy efficiencies are possible in nitride light emitters if the quantum well heterostructures are grown in the N-polar rather than the Ga-polar orientation. The reasons for this are discussed later – but at first glance it may seem that the polarity of the crystal heterostructure should have no effect on the IQE of the quantum wells. After all, if one considers just *photoluminescence* from the quantum wells in undoped structures, a N-polar quantum well heterostructure is equivalent to a Ga-polar heterostructure physically turned upside down, and indeed there is no reason to expect a difference in the IQE between the two structures. However, when a quantum well heterostructure is grown along the N-polar (000-1) direction of the crystal, it is found to have a drastically lower IQE than a structurally identical Ga-polar (0001) counterpart [4]. The reason for this remains a mystery, and unsolved to date. It is likely related to the difference in defect formation mechanics, e.g. higher layer contamination [5,6] for growths in the N-polar orientation by metalorganic vapor phase epitaxy (MOVPE) and by molecular beam epitaxy (MBE) [4,7-10], due to the drastically different growth dynamics and chemistry of the N-polar and Ga-polar structures.

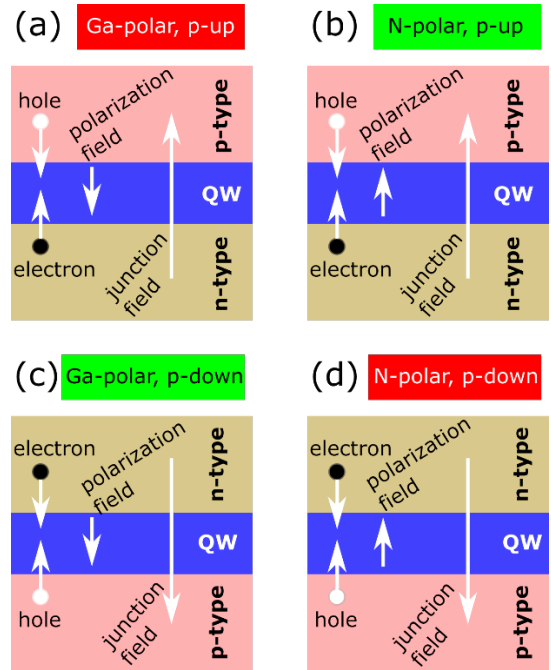

**Fig. 1** Schematic image showing four different structures realizing all possible combinations of GaN crystal polarization direction, and the choices of p-n junction. White arrows indicate electron and hole injection directions with respect to the built-in polarization fields and the doping-induced p-n junction fields.

If the reason for the low IQE of N-polar quantum wells is discovered and solved, and the IQE is brought up to the level of Ga-polar heterostructures, constructing LEDs (or laser diodes) requires placing the quantum wells inside the depletion region of a p-n junction diode. Now, the polarity of the diode (p-layer on the top or on the bottom) and the polarization of the crystal (Ga-polar or N-polar) combine to give four distinct possibilities, shown in **Fig. 1**. A Ga-polar structure can be p-up (conventional LED, Fig. 1a) or p-down (Fig. 1c), and a N-polar structure can be p-up (Fig. 1b) or p-down (Fig. 1d). We will show that among these the Ga-polar p-down (Fig. 1c) and N-polar p-up (Fig. 1b) structures are ideal for LEDs and LDs because the p-n junction field and the polarization fields are aligned. All of today's LEDs and LDs use Ga-polar p-up (Fig. 1a) which has some disadvantages. Because of the low IQE of N-polar quantum wells today, we select the Ga-polar p-down geometry (Fig. 1c), and experimentally prove our hypothesis that this structure has a) much brighter light emission, b) no carrier overflow nor undesired photon emission, and c) excellent current spreading.

To utilize the advantages of the higher perfection of growth along the Ga-polar [0001] direction along with the preferred polarization field orientation (available for p-up LEDs grown along the N-polar [000-1] direction), tunnel junctions (TJs) must be used on n-type substrates [11].  Growth of such structures is complicated due to the fact that the TJ must be grown *below* the quantum well active region.  This requires sustaining a high crystal quality and a smooth surface while growing highly doped p- and n-type layers. Additionally, growth of such structures by MOVPE is complicated, because success to effectively activate buried p-type layers is still limited [12,13] as the diffusion of hydrogen needed to activate the p-type Mg-doped GaN is forbidden through n-type GaN layers [14,15].  That is why to realize the bottom-tunnel junction, we have chosen to use plasma-assisted MBE, *a hydrogen-free growth technique* which does not require activation of the buried p-type layer. This growth technique has been proven capable of growing high quality nitride blue laser diodes [16].

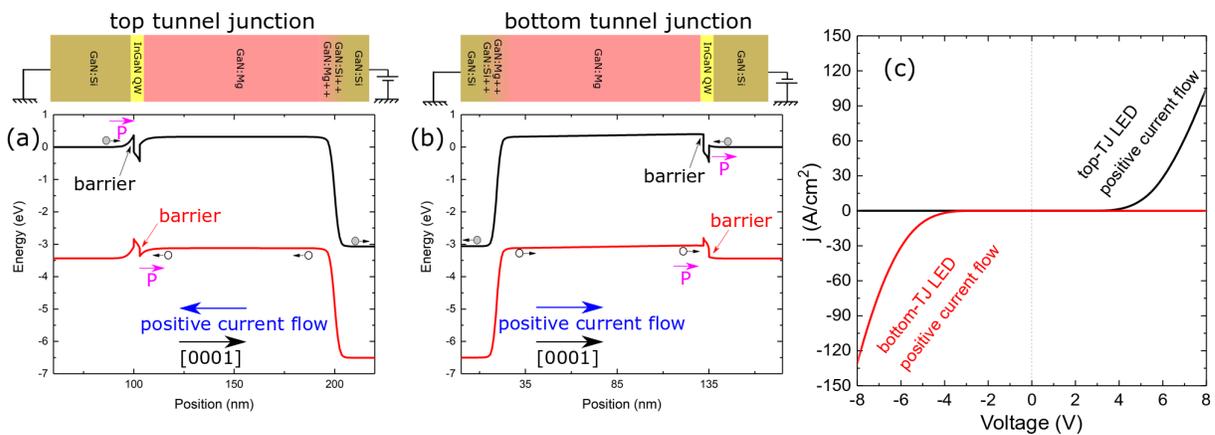

**Fig. 2** Schematic energy band diagrams for forward biased single QW LEDs utilizing top-TJ (a) and bottom-TJ (b) geometries, respectively. The growth is performed in the [0001] direction, proceeding from left to right.  Filled and empty circles with arrows denote carrier flow direction for electrons and holes, respectively.  Polarization in the vicinity of the quantum well is marked with respect to the current flow

and crystallographic [0001] direction. (c) Experimentally measured current densities as a function of external bias obtained for 80 μm x 80 μm LEDs utilizing the top-tunnel junction (a) and the bottom-tunnel junction (b).

In this paper, we analyze the impact of the direction of built-in electric field on the electrical and optical properties of MBE-grown LEDs. The desired heterostructures are realized in two varieties: by placing the TJ above the active region (top-TJ – shown in the energy band diagram in **Fig 2**(a)) and below the active region (bottom-TJ – shown in the energy band diagram in **Fig 2**(b)). Electron (filled circles) and hole (empty circles) flow directions are depicted inside the LEDs under forward bias conditions. The experimentally measured current densities for such structures as a function of the external bias is shown in **Fig 2**(c). As expected, the forward bias voltage direction for the bottom-TJ LEDs (as in Fig 2(b)) is *inverted* compared to that for the top-TJ LEDs (as in Fig 2(a)), indicating the *first successful realization* of a bottom-tunnel junction LED, which makes it possible to probe the questions raised earlier. Before discussing the electroluminescence properties, we describe the details of the epitaxial growth of the heterostructures and their fabrication into LEDs.

**Experimental**

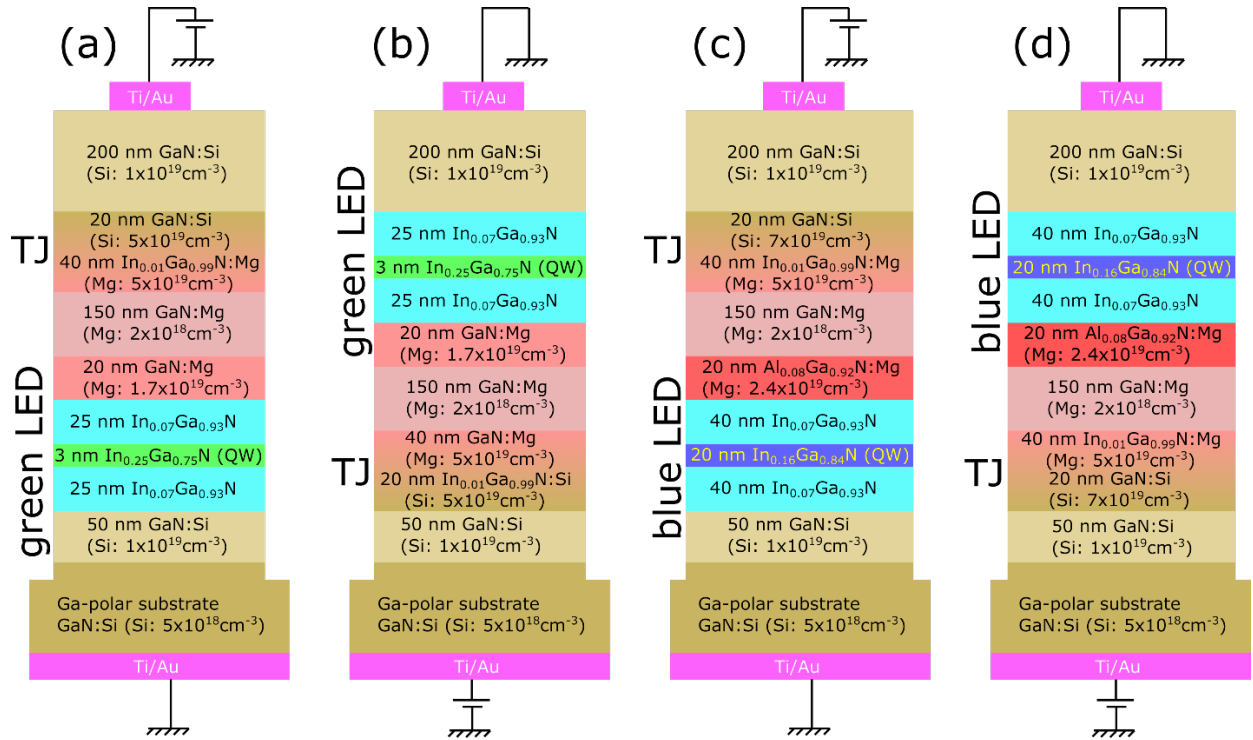

**Fig. 3** Layer details of the quantum well heterostructures and doping of LEDs grown for comparison of the impact of built-in polarization-induced electric field direction. All substrates are Ga-polar. (a, b) have a ~25% In content 3 nm thick InGaN quantum well for green light emission, and (c, d) have a 16% In content 20 nm thick InGaN quantum well for blue light emission. (a, c) have top-TJs with an opposite junction field direction to (b, d), which have bottom-TJs. Structures presented in (a), (b), (c) and (d) will be referred to as A, B, C and D, respectively. TJ and LED region are marked for clarification.

The samples shown in **Fig. 3** were grown by plasma-assisted MBE on commercially available bulk n-type Ga-polar GaN substrates with threading dislocation density around $5 \times 10^7$ cm$^{-2}$. The composition of the active InGaN layers are chosen for two wavelengths: blue and green, and for each choice, a tunnel junction is inserted on the top and the bottom. The top TJ structures are similar to those in prior studies [17,18] and serve as control samples, while the bottom-TJ structures are novel. The InGaN layers were grown at

650°C and the GaN layers at 740°C. In each series, the samples were grown using exactly the same doping concentrations and the same alloy compositions. Modulation of In and Mg concentration for layers grown at the same temperature was achieved by varying atomic fluxes of Ga [19] and Mg [20], respectively. To assure the same growth temperature for all equivalent layers, metal-desorption time measurements of reference In and Ga fluxes were performed before the growth. The surface roughness measured by atomic force microscopy for 5 μm x 5 μm regions for all structures after MBE growth was below 0.5 nm indicating a smooth morphology.

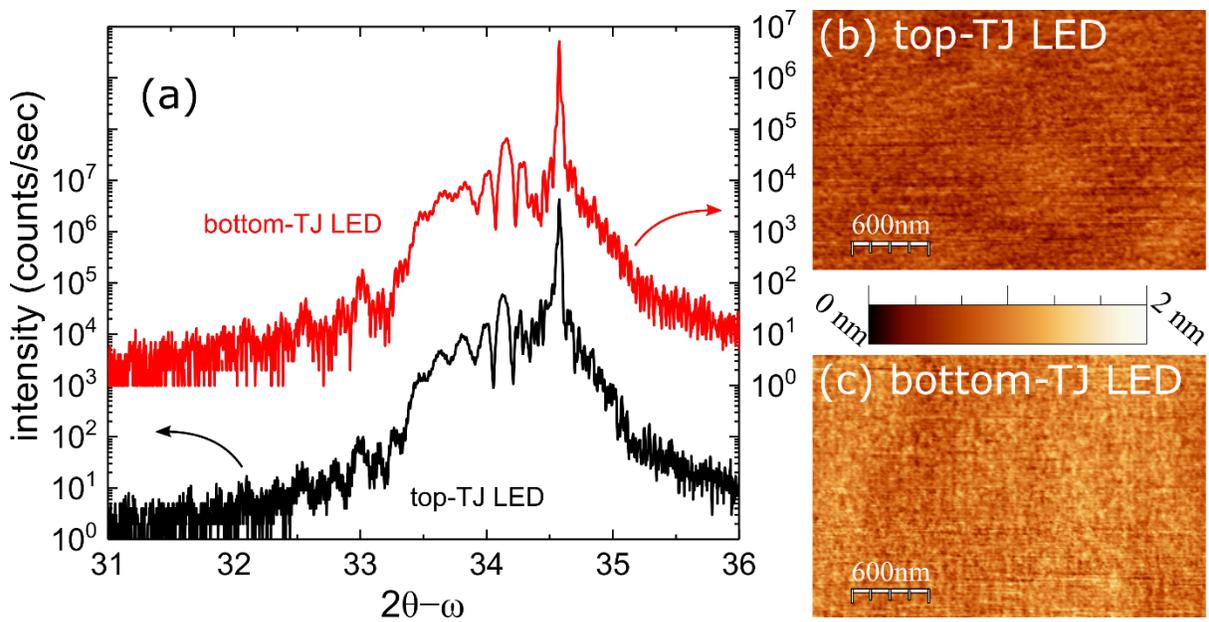

**Fig. 4** (a) X-ray diffraction 2θ-ω scans obtained for structures C and D on left and right axes, respectively. The placement of the doping-induced tunnel junction on the top or the bottom has no discernible effect on the X-ray diffraction pattern. In (b,c) AFM images obtained for samples C and D, respectively, are presented using the same scale. Root mean square roughness obtained for 2 μm x 3 μm scans for both structures was 0.15 nm and 0.2 nm for top-TJ (b) and bottom-TJ (c) LEDs, respectively.

In **Fig. 4**, high-resolution X-ray diffraction 2θ-ω scans and AFM images measured for samples in Fig 3 (c, d) are shown. They indicate successful growth of essentially identical chemical and layer compositions and thicknesses for both the top-TJ and bottom-TJ heterostructures, and simultaneously prove that a) the doping sequence has no impact on the X-ray diffraction patterns and samples morphologies and b) the placement of the TJ at the bottom does not adversely affect the compositional control of the subsequently grown layers.  As shown in Fig 3 (a, b) (structures A and B) and (c, d) (structures C and D), the difference between the two series is in the active region thickness and composition of the quantum well (QW) and barriers, composition of the electron blocking layer (EBL), and slight difference in doping levels in the TJs. The InGaN layers surrounding the active regions are used to keep the structure similar to laser-diode heterostructure design used in our prior work [16].  Relatively thick QWs in structures C and D represent a new design of light emitters enabled by the TJs.  We also note that unlike polarization-induced heterostructure tunnel junctions [11,14,21-25], we have chosen to use GaN:Si/GaN:Mg homojunction TJs to avoid potential complications of growing active regions on top of buried heterostructures.  Single quantum well active regions were used to remove the effect of uneven carrier distribution between wells for both the blue and green LED structures to obtain a fair assessment of the impact of the built-in electric field on the intrinsic performance of LEDs for both polarities.  The bottom-TJ effectively flips the LED design upside down, mimicking a N-polar p-up configuration, but in the Ga-polar crystal growth orientation.

The samples were processed using optical lithography into vertical diodes of different die sizes ranging from 20 μm x 20 μm to 300 μm x 300 μm.  Several higher aspect ratio dies were also fabricated in order to observe current spreading, with the largest aspect ratio device measuring 100 μm x 500 μm.  ICP-RIE mesa etching in Chlorine based chemistry was used to isolate individual devices.  Though not exploited in this work, we point out that the use of bottom-TJ structures offers a *unique advantage* of lowering the p+/n+ tunneling contact resistance by exploiting a much larger cross-section area than the LED mesa, something that is not possible in the top-TJ LED structure. After mesa isolation, a top circular Ti/Al ohmic contact of radius 25 μm for the 80 μm x 80 μm and larger devices was deposited at the center of the top of the mesas. Finally, a common bottom contact was deposited on the back of the n-type substrate. Prior to metallization, a surface treatment consisting of $O_2$ plasma ashing, HCl, and HF cleans were performed to remove undesired chemical residues. The same 25 nm Ti/100 nm Al metal stack was deposited by e-beam evaporation for both the top and bottom n-type contacts. Such a contact stack has been shown to have very low contact resistance to n-type GaN [26].

From this point on, top and bottom TJ samples will be compared using opposite bias directions (as indicated in **Fig 3**) to enable quantitative comparison in their natural bias conditions. The processed samples were measured on-chip by grounding the bottom contact and putting a probe on the top contact of the selected devices. The electroluminescence (EL) spectrum was collected from the top using an optical setup utilizing a monochromator and a photodiode.

**Results**

The measured I-V characteristics of the four 80 μm x 80 μm devices are shown in **Fig. 5**. As expected from the comparable doping levels, current values obtained for bottom- and top-TJs are at a similar level with higher values for bottom-TJ structures near the turn-on voltage (easily seen in the inset of the **Fig 5**(a)).

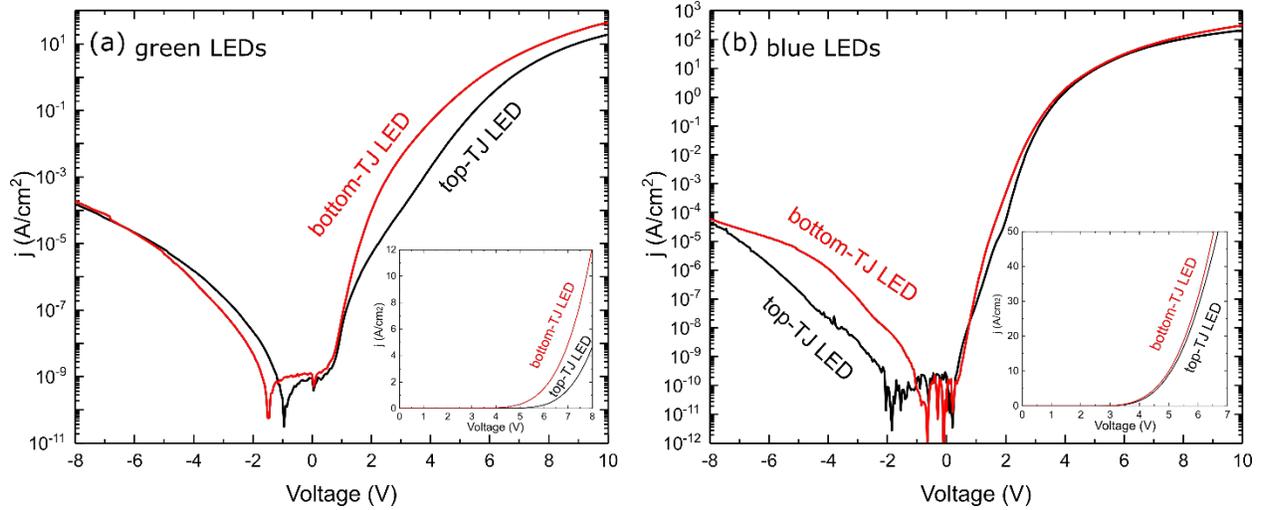

**Fig. 5** Measured current density versus voltage for 80 μm x 80 μm devices from green LED structures (a) with top TJ A and bottom TJ B, and blue LED structures with (b) top-TJ C and bottom-TJ D. The insets show the same plot in linear scale in the forward bias. The measurements show low leakage currents and the successful realization of bottom-TJ heterostructures for both the green and blue LEDs.

The low observed leakage levels in the reverse bias for all LED structures seen in **Fig 5** importantly indicates that the density of extended defects propagating through the LEDs in all cases is similar. This confirms high epitaxial quality of bottom-TJ structures, and thereby enables a fair comparison between the top-TJ and bottom-TJ structures. The lower turn-on voltages seen in the insets of **Fig 5** for the blue LEDs whose structures are shown in **Fig. 3 (c,d)** are due to a slight increase in the doping level in the 20 nm thick GaN:Si TJ region from $5 \times 10^{19}$ to $7 \times 10^{19}$ cm$^{-3}$. It is important to point out that the on-resistances can be further reduced by increasing Si and Mg doping, and by potentially using polarization-induced TJs with introduction of InGaN or AlN layers in the TJ [21-24,27], but with a potential penalty of degraded crystal quality in the active region.

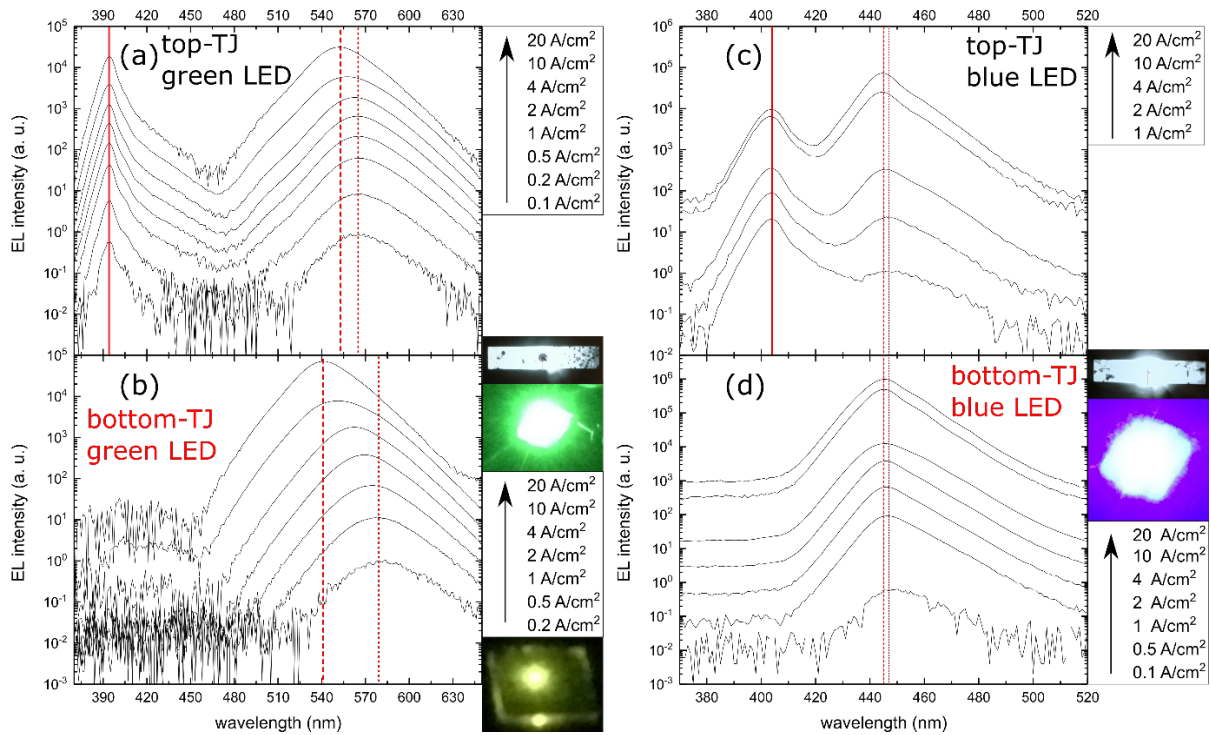

**Fig. 6** (a-d) Electroluminescence spectra in log scale measured on-chip for the indicated current densities for 80 μm x 80 μm devices of structures A, B, C and D, respectively. A large energy emission shift is observed in (b), accompanied by real-color pictures of the whole 1 cm x 1 cm wafers next to the plot. No visible peak shift was observed for structures C and D. Above the real-color images are monochromatic images collected under a microscope for 100 μm x 500 μm bottom-TJ devices under 100 mA injection, showing excellent current spreading. The bottom-TJ structures B and D show single peak emission, whereas both blue and green top TJ structures show parasitic peaks at a higher photon energy than the desired quantum well peaks.

**Fig. 6** (a-d) shows the measured electroluminescence spectra for the four tunnel-junction LEDs. The dotted and dashed vertical lines indicate the peak emission wavelength emitted by the quantum wells at low and high excitation, respectively. The solid vertical line indicates EL associated with parasitic radiative recombination in the lower indium content InGaN layers surrounding the quantum wells. Rather interestingly, such a high energy emission peak (395-405 nm) appears for all top-TJ LEDs but *never* for the bottom-TJ LEDs. Measurements performed on different devices on the same sample (for top-TJ) feature slight differences in relative intensities, but existence of the secondary peak and its dominant intensity at low current density is observed for *all* top-TJ LEDs. It is important to point out that this extra high energy peak is *not* associated with emission or reemission from the p-type doped GaN layers, as it is also measured in photoluminescence measurements of similar InGaN/InGaN QWs [28] in the *absence* of Mg doped p-type GaN layers.

Due to the limitations of on-chip measurements, an exact comparison of the intensity between the top- and bottom-TJ diodes is not straightforward due to differences in alignment of probes, etc. Nonetheless, at an injection current density of at 20 A/cm$^2$, the bottom-TJ LEDs showed ~2.5X higher peak intensity than the top-TJ for the green LEDs, and this enhancement was ~13X for the bottom-TJ blue LEDs. The quantitative differences between the enhancement for green and blue emitters can be attributed to differences in active regions and electron blocking layer (EBL) design, but irrespective of the details, the bottom-TJ structures for both wavelengths demonstrate the important advantages offered by this conceptual change in the LED design. For the higher In content LEDs (**Fig 6**(a,b)), the peak emission shifts from yellow (top-TJ: 565 nm, bottom-TJ: 580 nm) to green (top-TJ: 552 nm, bottom-TJ: 541 nm) with increasing current injection. A higher blueshift for the bottom-TJ device is seen compared to the top-TJ: this is a fingerprint of higher injection efficiency at high currents, leading to more pronounced screening of the internal polarization field. This is equivalent to say that at the same current level, the effective carrier concentration in the InGaN QW is higher in the bottom-TJ LEDs than the top-TJ LEDs, thus more screening and more light emission.

Irrespective of the location of the TJ, all LED devices showed good current spreading: the small circular metal contact on the top was enough to spread the current throughout the entire device mesas far larger in area. Optical microscope images of the highest aspect ratio (100 μm x 500 μm) bottom-TJ LEDs under 100 mA injection are shown next to the EL spectra measurements in **Fig 6**.

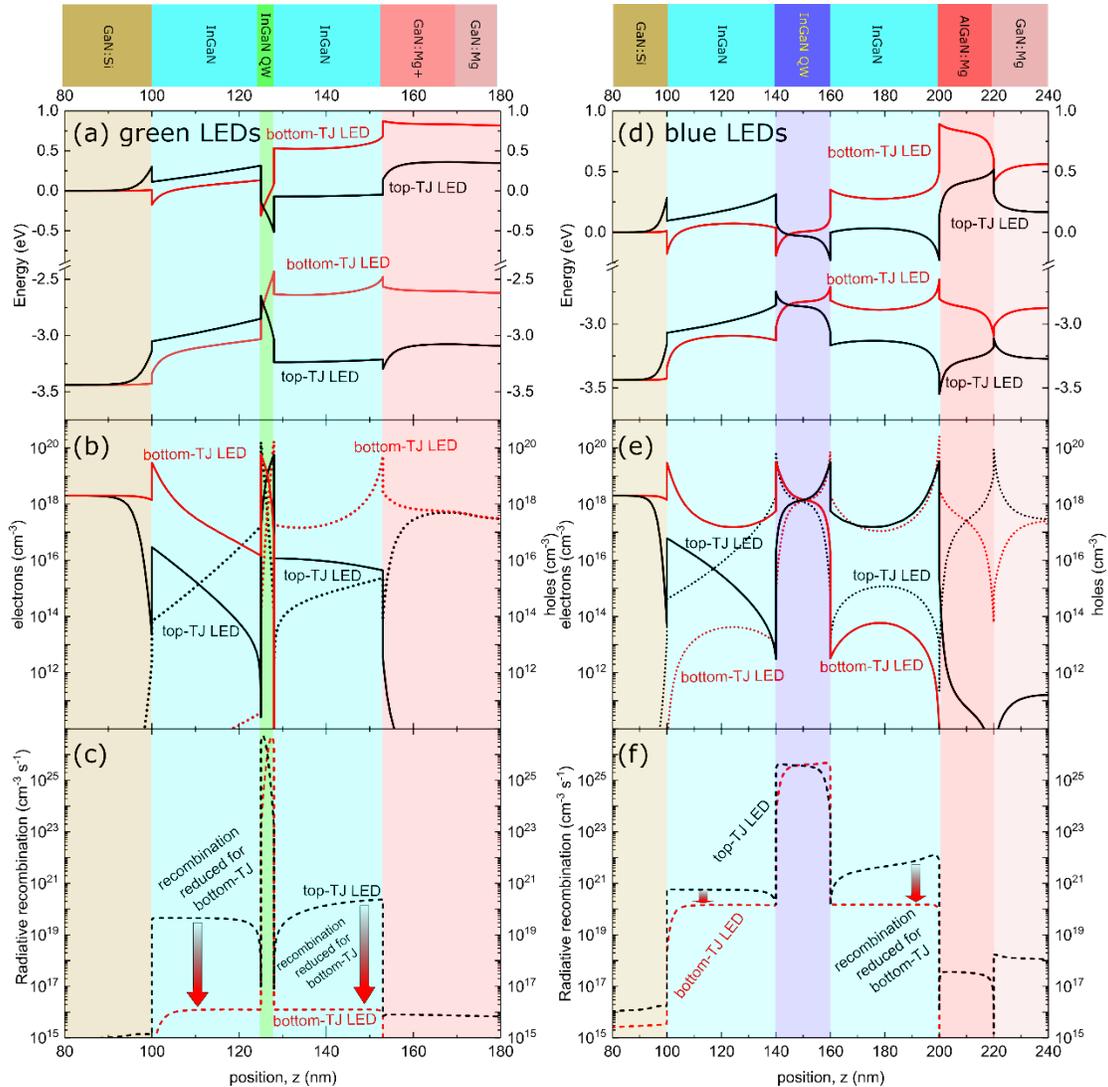

**Fig. 7** Energy band diagrams, mobile carrier concentrations, and radiative recombination as a function of position near the active region calculated by the SiLENSe package (a,b,c) at 20 A/cm$^2$ for structures A and B and (d,e,f) at 20 A/cm$^2$ for structures C and D. The results for top-TJ and bottom-TJ structures are compared. (b,e) Electron and hole concentrations are denoted using solid and dotted lines, respectively. (c,f) Differences in the radiative recombination in the InGaN claddings surrounding the QW between top- and bottom-TJ structures are marked with arrows.

To further analyze the origin of the high energy peak in the EL spectra for top-TJ structures, the heterostructure devices were modeled by self-consistently solving the energy band diagrams with current transport through the structure. The experimentally investigated structures utilizing top- and bottom-TJ were simulated by one-dimensional drift-diffusion model using SiLENSe: only the part of the structures marked `green LED' or `blue LED' in **Fig 3** (excluding the part denoted "TJ") were simulated, while setting the orientation as Ga- and N-polar, respectively. This procedure works well for modeling physical differences between the top- and bottom-TJ LEDs as long as a) the TJ resistance is low in both cases, b) the TJ is not affected by the polarization field direction (e.g. by using a homojunction), and c) does not degrade the LED active region quality.

The calculated energy-band diagrams, carrier concentrations and radiative recombination profiles are shown in **Fig 7**. These plots present results obtained at 20 A/cm$^2$ for the green LED structures A (top TJ) and B (bottom TJ) and 20 A/cm$^2$ for the blue LED structures C (top TJ) and D (bottom TJ). As seen in the energy band diagrams, the top-TJ and bottom-TJ LEDs behave quite differently under forward bias. In the top-TJ, the built-in polarization field in the QW lowers the effective band offset for electrons and holes, enabling the carriers to exit the QW and overshoot to the other side. This increases the escape of carriers from QWs, leading to higher carrier concentrations outside of the QW for the top-TJ structure, compared to the abrupt drop of carrier overshoot for the bottom TJ structure, as seen in **Fig 7**(b,e). As a consequence, a significant amount of *undesired* radiative recombination occurs in the barriers surrounding the QW for the top TJ structures (**Fig 7**(c,f)). For the bottom-TJ LEDs on the other hand, due to the inverted polarization field direction in vicinity of the QW, electrons and holes are retained in the QW and in the InGaN layer closer to the desired injection side, respectively, as seen in **Fig 7**(b,e). Because of the reduction of the n*p product outside the QWs, the recombination in the barrier surrounding the QW is significantly reduced for the bottom-TJ LEDs compared to the top-TJ case (**Fig 7**(c,f)). The parasitic emission peaks seen in **Fig 6**(a,c) in only the top-TJ LEDs are exactly due to this recombination in the low In-composition InGaN cladding regions, and are absent in the bottom-TJ LEDs.

This change in the radiative recombination rate between top- and bottom-TJ LEDs depends on the composition and thickness of the QW, barriers, and the EBL, which can be inferred from comparing **Fig 7**(c) and **Fig 7**(f). The simulations suggest that the bottom-TJ construction always leads to significantly *lower* undesired radiative recombination rates in the barriers. It is important to point out that in **Fig 7**(c,f), only the radiative part of recombination is plotted to relate those results with the experimentally obtained high energy peak in the EL. An important consequence of the efficient electron and hole concentration separation in the vicinity of QW is the resulting lowering of nonradiative recombination, which is also proportional to the n*p product. This fact significantly enhances the efficiency of the bottom-TJ LEDs.

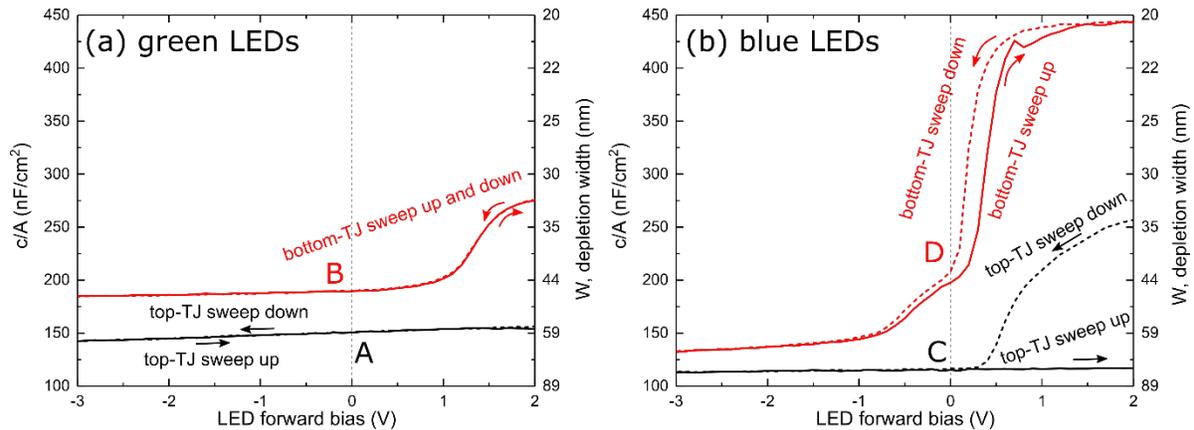

**Fig. 8** Capacitance-Voltage (C-V) measurements corresponding to structures (a) A and B, and (b) C and D. The higher junction capacitances as well as the minimal hysteresis observed in the bottom-TJ LEDs arise from the polarization field in the QW pointing in the same direction with the p-n junction field, which are further illustrated in Fig. 9. The CV measurements were taken at 5 MHz.

The low leakage current in reverse bias seen in **Fig 5** allows a direct measurement of the capacitance-voltage (C-V) characteristics for all TJ LED samples. In Fig **8**(a,b), the capacitance values divided by device area ($C/A$) measured at a 5 MHz frequency as a function of forward bias of the LEDs are shown. The plots in **Fig 8**(a) and (b) present data for the structures A, B and C, D, respectively. The depletion width $W$ corresponding to the measured $C/A$ values, assuming $W = \frac{\varepsilon_r \varepsilon_0 A}{c}$, where $\varepsilon_0 = 8.85 \cdot 10^{-8} \mu F/cm$ and $\varepsilon_r = 10$ (the electrical permittivity of vacuum and the semiconductor dielectric constant) are shown on the right vertical axes. For simplicity, $\varepsilon_r$ is assumed to be constant for $In_{0.15}Ga_{0.85}N$ and GaN – derived from a linear extrapolation proportional to the In content using values of 8.52 and 18.44 for GaN and InN, respectively [29]. Since the conductivity of the TJs is high, we assume that the entire measured capacitance is associated with the depletion in the LED active region. Due to the opposite orientations of the polarization-induced electric field and the p-n junction electric fields of the p-up (top TJ) and p-down (bottom-TJ) LEDs, the chemically identical layer stacks result in remarkably different capacitances and depletion widths. As expected, $W$ obtained for the bottom-TJ structures is significantly *lower* than that for top-TJ structures, implying the polarization field is aligned with the junction field, and to maintain the same net built-in field, the junction shrinks the net depletion width (see **Fig 1b**). This mimics a desired N-polar p-up structure (see **Fig 1c**) in the alignment of the polarization and junction fields, but by virtue of the bottom-TJ, is effectively realized in the Ga-polar structure. For the bottom-TJ LEDs, a characteristic steep increase in the capacitance is measured for positive voltages, associated with barrier-free carrier injection and ready shrinking of the depletion regions. For example, in structure D at a forward bias voltage of ~1 V, the depletion width measured from the capacitance is essentially identical to the thickness of the QW, ~20 nm, meaning the depletion region has shrunk completely to the edges of the QW. On the other hand, a noticeable hysteresis is observed between forward and reverse voltage sweeps for the top-TJ LED, associated with charge trapping in the active region. This effect is smaller for thinner QWs and for bottom-TJ structures compared to top-TJ structures. For voltages higher than 1 to 2 V, the onset of current flow makes the conductance dominate over the capacitance, and the C-V measurement should not be trusted.

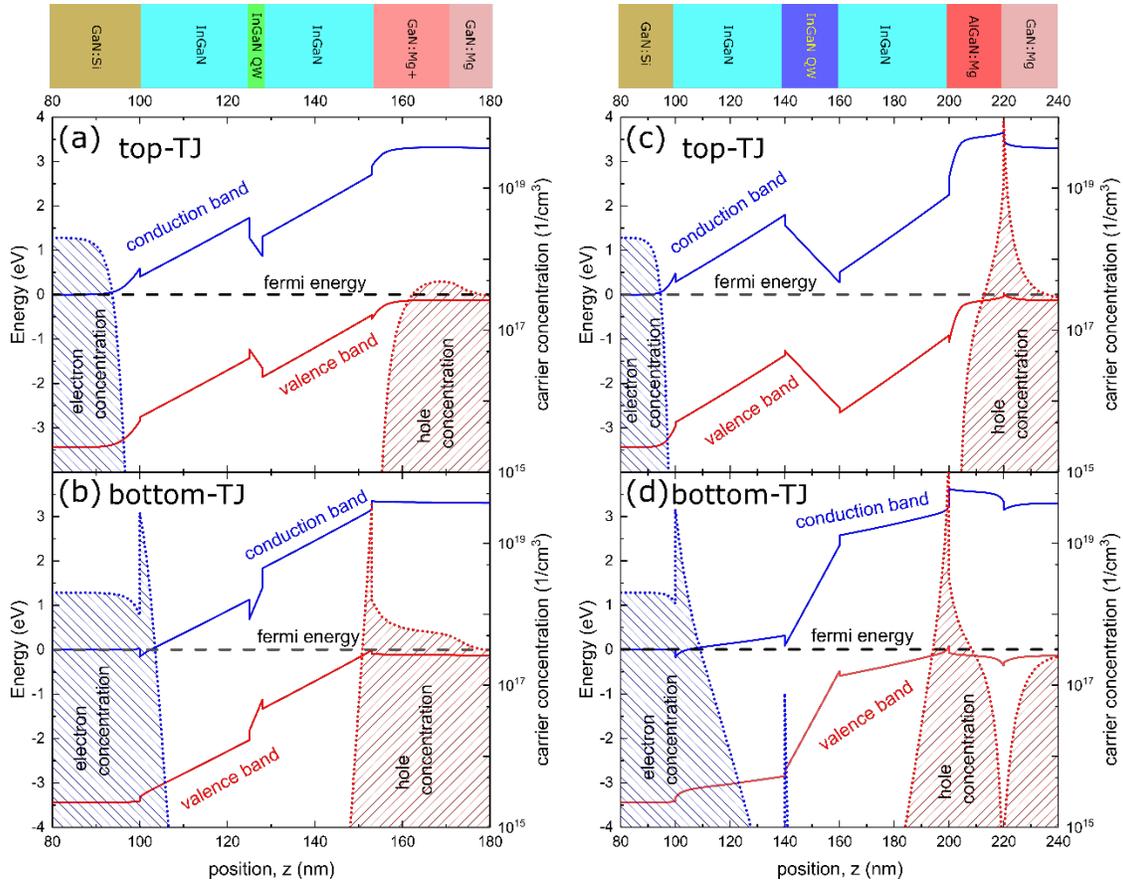

**Fig. 9** Energy band diagrams with carrier concentration at 0 V obtained using the SiLENSe package (a,b,c,d) comparing results obtained for structures A, B, and C, D. The top-TJ blue LED has a triangular well at 140-160 nm that can trap charges. The bottom-TJ green and blue LEDs have smaller depletion regions.

To benchmark the magnitude of the experimentally obtained depletion widths, energy band diagram simulations using SiLENSe at 0 V were performed. The calculated energy band diagrams and mobile carrier concentrations in the vicinity of the active regions of structures A, B, C and D are shown in **Fig 9**(a-d), respectively. It is seen clearly that the polarization-induced electric field results in the thinning of the net depletion width for the bottom-TJ structures vs top TJ structures (seen for B vs A and D vs C). The peaks in the electron and hole concentrations near the active region that are present for the bottom-TJ LED structures are caused by the GaN/InGaN and InGaN/AlGaN heterojunction band offsets, and promote better carrier injection into QW. The AlGaN/GaN interface that causes the peak in the hole concentration for structure C is on the other side of the EBL, leading to lower hole injection efficiency. The energy band diagrams also provide insight into the origins of the hysteresis observed in C-V for sample C. A deep triangular well is formed around 160 nm in the conduction band, with barriers on both sides. The charging/discharging of this well formed around the InGaN QW due to polarization due to carriers injected or extracted from this region is responsible for hysteresis. The same barrier is responsible for blocking carrier injection into the QW at low currents, leading to a dominant parasitic high energy peak associated with emission from InGaN cladding layers in the EL spectra of **Fig 6**(c). In the case of structure D, which also has the thick QW, the barrier for carrier escape is significantly smaller, leading to significantly lower hysteresis as seen in **Fig 8**(b) for the BTJ vs the top TJ. Such carrier trapping in the top TJ LEDs can be detrimental for fast electronic modulation of the LEDs or LDs, suggesting the bottom-TJ construction is superior for such high-speed applications as in visual light communications (VLC) or LiFi (Light-fidelity).

**Conclusions**

In this work, a fundamentally new approach of using bottom-tunnel junctions for polar GaN based light-emitting diodes and lasers has been demonstrated. A clear experimental assessment of the impact of the polarization-induced electric fields and their relative orientation with the p-n diode junction fields on LED performance was possible for the first time with the bottom-TJ approach. Electroluminescence measurements on top-TJ and bottom-TJ LEDs emitting in blue as well as green-yellow spectrum indicate that the bottom-TJ LEDs achieve superior injection efficiency (in a Ga-polar p-down structure, which effectively mimics a desired N-polar p-up LED structure). Corroborating device simulations explain the experimental trends observed in the EL spectra such as higher emission intensities at high injection currents for bottom-TJ over top-TJ structures: the polarization induced electric fields present in the bottom-TJ structures are in a direction that assists carrier injection, while effectively separating electrons and holes in the vicinity of the QW, suppressing parasitic carrier recombination. Further supporting capacitance-voltage measurements confirm the experimental observation of significantly thinner depletion widths and better injection efficiency for the bottom-TJ LEDs compared to the top-TJ counterparts. Based on the comparative study, we conclude that the bottom-TJ LEDs present an intriguing alternative to standard Ga-polar LEDs for achieving high efficiencies, monochromaticity, and for high frequency direct modulation of the photonic output. Combining the above advantages with the excellent current spreading by using a thick n-type GaN layer on the top, the first realization of a bottom-tunnel junction GaN LED presented here paves a new unexplored path for GaN-based photonic devices.

**Acknowledgements**

This work was supported partially by the Polish National Centre for Research and Development grants LIDER/287/L-6/14/NCBR/2015 and PBS3/A3/23/2015 during the visit of the lead author at Cornell University. The work at Cornell University was supported in part by the following National Science Foundation (NSF) grants: NSF DMREF award # 1534303 monitored by Dr. J. Schluter, NSF Award # 1710298 monitored by Dr. T. Paskova, NSF CCMR MRSEC Award #1719875, and NSF RAISE TAQs Award #1839196 monitored by Dr. D. Dagenais. The cleanroom fabrication at the Cornell Nanofabrication Facility (CNF) was supported in part by the NSF National Nanotechnology Coordinated Infrastructure (Grant ECCS-1542081).